\documentclass[sigconf]{acmart} 
\AtBeginDocument{%
  \providecommand\BibTeX{{%
    \normalfont B\kern-0.5em{\scshape i\kern-0.25em b}\kern-0.8em\TeX}}}

\copyrightyear{2024}
\acmYear{2024}
\setcopyright{acmlicensed}\acmConference[CHI '24]{Proceedings of the CHI Conference on Human Factors in Computing Systems}{May 11--16, 2024}{Honolulu, HI, USA}
\acmBooktitle{Proceedings of the CHI Conference on Human Factors in Computing Systems (CHI '24), May 11--16, 2024, Honolulu, HI, USA}
\acmDOI{10.1145/3613904.3642287}
\acmISBN{979-8-4007-0330-0/24/05}

\usepackage{amsmath}  

\usepackage{enumitem}

\newlist{oddenumerate}{enumerate}{1}
\setlist[oddenumerate]{start=0,label=\theoddenumeratei.}
\makeatletter
\renewcommand\theoddenumeratei{\@arabic{\numexpr2*\value{oddenumeratei}+1}}
\makeatother

\newlist{evenenumerate}{enumerate}{1}
\setlist[evenenumerate]{start=0,label=\theevenenumeratei.}
\makeatletter
\renewcommand\theevenenumeratei{\@arabic{\numexpr2*\value{evenenumeratei}+2}}
\makeatother

\begin{document}


\title[Signmaku: Sign Language Danmaku]{Towards Inclusive Video Commenting: Introducing Signmaku for the Deaf and Hard-of-Hearing}




\author{Si Chen}
\email{sic3@illiois.edu}
\orcid{0000-0002-0640-6883}
\affiliation{%
  \institution{School of Information Sciences, University of Illinois Urbana-Champaign}
  \city{Champaign}
  \state{Illinois}
  \country{USA}
  \postcode{61802}
}

\author{Haocong Cheng}
\email{haocong2@illinois.edu}
\affiliation{
  \institution{School of Information Sciences, University of Illinois Urbana-Champaign}
  \city{Champaign}
  \state{Illinois}
  \country{USA}
}

\author{Jason Situ}
\email{junsitu2@illinois.edu}
\affiliation{
  \institution{Computer Science, University of Illinois Urbana-Champaign}
  \city{Urbana}
  \state{Illinois}
  \country{USA}
}

\author{Desirée Kirst}
\email{deskirst@gmail.com}
\affiliation{
  \institution{Gallaudet University}
  \city{Washington}
  \state{District of Columbia}
  \country{USA}
}

\author{Suzy Su}
\email{xiaoyus4@illinois.edu}
\affiliation{
  \institution{School of Information Sciences, University of Illinois Urbana-Champaign}
  \city{Champaign}
  \state{Illinois}
  \country{USA}
}

\author{Saumya Malhotra}
\email{saumyam2@illinois.edu}
\affiliation{
  \institution{College of Liberal Arts \& Sciences, University of Illinois Urbana-Champaign}
  \city{Urbana}
  \state{Illinois}
  \country{USA}
}

\author{Lawrence Angrave}
\email{angrave@illinois.edu}
\affiliation{
  \institution{Computer Science, University of Illinois Urbana-Champaign}
  \city{Urbana}
  \state{Illinois}
  \country{USA}
}

\author{Qi Wang}
\email{qi.wang@gallaudet.edu}
\affiliation{
  \institution{Gallaudet University}
  \city{Washington}
  \state{District of Columbia}
  \country{USA}
}

\author{Yun Huang}
\email{yunhuang@illinois.edu}
\affiliation{
  \institution{School of Information Sciences, University of Illinois Urbana-Champaign}
  \city{Champaign}
  \state{Illinois}
  \country{USA}
}

\begin{abstract}

Previous research underscored the potential of danmaku--a text-based commenting feature on videos--in engaging hearing audiences. Yet, for many Deaf and hard-of-hearing (DHH) individuals, American Sign Language (ASL) takes precedence over English. To improve inclusivity, we introduce ``Signmaku,'' a new commenting mechanism that uses ASL, serving as a sign language counterpart to danmaku. Through a need-finding study (N=12) and a within-subject experiment (N=20), we evaluated three design styles: real human faces, cartoon-like figures, and robotic representations. The results showed that cartoon-like signmaku not only entertained but also encouraged participants to create and share ASL comments, with fewer privacy concerns compared to the other designs. Conversely, the robotic representations faced challenges in accurately depicting hand movements and facial expressions, resulting in higher cognitive demands on users. Signmaku featuring real human faces elicited the lowest cognitive load and was the most comprehensible among all three types. Our findings offered novel design implications for leveraging generative AI to create signmaku comments, enriching co-learning experiences for DHH individuals.

\end{abstract}



\maketitle

\section{Introduction}

``Danmaku'' is a video-commenting feature where users overlay their \textit{text} comments directly onto videos; its effect on learning lies in fostering a more interactive and engaging learning environment by allowing learners to view comments from their peers, encouraging active participation and social connection during the video-based learning process \cite{wu2019danmaku}. However, text is less accessible for Deaf and hard-of-hearing (DHH) users, as many of them prefer signed language over text, in our context American Sign Language (ASL) over English \cite{mack2020social}. More than 500,000 people in the United States use American Sign Language (ASL) as their primary mode of communication \cite{articlele}.

Existing research focuses on improving the accessibility of video-based learning for DHH populations via generating closed captions, and text versions of the audio content. Despite accounting for reading proficiency, the differences in background knowledge and information processing strategies make text-based information less accessible for DHH than hearing \cite{marschark2005access}. On the other hand, sign language is a form of visual communication that employs gestures, facial expressions, and body language. It is a first-class language with the same level of complexity as spoken English, but it uses a different sentence structure \cite{rastgoo2021sign}. Little is known about how DHH users perceive and share comments in ASL.

\begin{figure*}[ht]
    \includegraphics[width=0.9\textwidth]{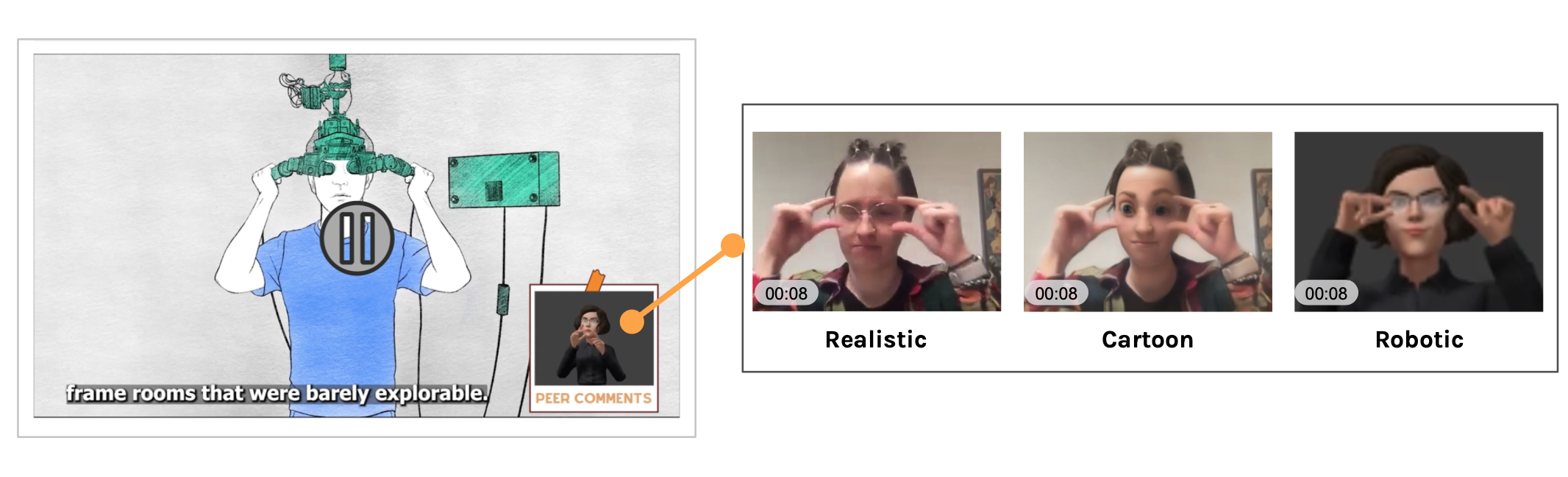}
    \caption{Three Styles of Signmaku designs explored in own study: \textit{realistic} (unfiltered), \textit{cartoon} (filtered face with real torso and hands), and \textit{robotic} (filtered face, torso, and hands) leverages AI technology to filter signed video clips, considering different styles of privacy preservation. \textit{Cartoon} filter's background was selected to contrast an individual's skin color for more visible hand movements using VToonify \cite{zhang2022s}. \textit{Robotic} filter's appearance was customized by the DHH individual in DeepMotion.} 
    \label{fig:example}
    \Description{The left image is a screenshot of a video being played, with ASL-based comments embedded (signmaku) in the lower-right corner and open captions on the bottom not overlapping the signmaku. The right image shows the three types of Signmaku designs: realistic, cartoon, and robotic. The realistic design has no filter applied to the Signmaku. The cartoon design has a filter with a cartoon-like face, a real torso, and a real hand. The robotic design has a filter with robotic faces, torso, and hands.}
  \end{figure*}

Thus, in this paper, we introduce a new ASL-based commenting feature called Signmaku, which is like a sign language version of danmaku. Unlike text-based comments, signmaku offers ASL-based comments that contain visual information with facial expressions and hand movements, as shown in Fig.~\ref{fig:example}. The unique challenge of sign language, compared to text, is its heightened privacy concerns of revealing individuals' faces. Therefore, we designed three visual styles by applying different generative AI techniques: 1) \textit{Realistic} (unfiltered or original ASL recording with users' real faces); 2) \textit{Cartoon} (filtered faces with real torso and hands using VToonify \cite{zhang2022s}); and 3) \textit{Robotic} (filtered faces, torsos, and hands using DeepMotion\footnote{\url{https://www.deepmotion.com}}) that address different levels of privacy concerns. We aim to explore how it can improve the inclusivity of video-based learning for DHH. While our current signmaku design could not fully replicate the experience with text-based danmaku, we expect signmaku to enrich the learning experience of DHH students by allowing them to make and view comments for and from their peers in sign language, fostering a pseudo-synchronous co-learning environment.


In order to design and explore the effect of signmaku on DHH's video-based learning, we completed a two-phase study. In Phase I, we conducted a need-finding study with 12 DHH participants, which allowed us to identify several key requirements of the signmaku feature and collect high-quality signmaku comments that were used in the next phase. In Phase II, we generated differently styled signmakus and designed a within-subject experiment with 20 DHH participants, during which they viewed content with the collected signmakus while watching a video.  Participants were also asked to create and share their own Signmaku comments and choose their preferred style for sharing.  The design aim for the two filtered styles is to provide options for learners to remain anonymous if they so choose, and thereby eliciting their willingness to make and share their own ASL comments. The experimental study addresses: 
\begin{itemize}
    \item \textbf{RQ1}: \textit{How does viewing three styles of signmaku impact DHH learners' video-based learning?} 
    \item \textbf{RQ2}: \textit{How do DHH learners perceive the creation and sharing of the three styled signmaku? }

\end{itemize}



Our research makes significant contributions: 1) we propose a novel interaction design, Signmaku, which not only improves the inclusivity of video-based learning for DHH but also enhances their sense of social connectedness; 2) we provide empirical evidence of the effectiveness of cartoon-styled signmaku on learners' improved engagement and their reduced efforts of sharing comments with their peers, with fewer privacy concerns than the other two signmaku styles; and 3) we reinforce the importance of reproducing movements of certain facial areas for delivering meaning and emotions of DHH users with sign language, which is often overlooked by generative AI technologies. Our study primarily focuses on interaction design and understanding user behavior and does not contribute to technical advancements.

\section{Related Work}





\subsection{Challenges Faced by DHH Students in Video-based Learning}


Video-based Learning is a remote learning approach that relies on live or prerecorded video to teach new skills and knowledge. Many video-based learning platforms have attracted hundreds of thousands of students, such as MOOC, and YouTube. They allow people from around the globe to participate in courses via videos. For DHH, video-based learning is not yet fully accessible. The major reason is because of online videos in general lack good-quality captions, if any. Some research shows DHH people used filtering tools to find videos with captions for them to watch \cite{li2022exploration}. Even when captions exist, many are erroneous as they are auto-generated from automatic speech recognition algorithms \cite{bhavya2022exploring}. It is not unusual for these algorithms to miss some words due to noise and to recognize the incorrect word when generating captions \cite{kafle2016effect, berke2020deaf}. Human efforts are needed to correct those errors however video content creators were frequently unwilling or unable to do so \cite{li2022exploration, mcdonnell2022understanding}. Beyond the caption's accuracy, captions' appearance is also important to understanding \cite{berke2019preferred}. Low-quality captions for online learning videos created great challenges for DHH learners to understand the learning content when studying remotely from school through video lectures during the COVID-19 period, and aggravated mental health issues \cite{aljedaani2022if}. Furthermore, the United Nations acknowledged that COVID-19 exacerbates inequalities between people with and without disabilities, including access to education and inclusion in the community \cite{UnitedNations_2022}.

Social interaction encouraged within a video-based learning platform can help students stay engaged and enhance their own learning \cite{chen2021afraid, sablic2021video, yoon2021video}. Social engagement with peers is paramount for all learners, but it holds even greater significance for DHH students. This importance is magnified by the fact that a considerable number of DHH individuals have experienced or perceived exclusion within mainstream education (Mainstream is the practice of placing students with special education needs in a general education classroom.) 
Research indicates that deaf children can have social difficulties compared with their hearing peers \cite{stinson1999considerations, constantinou2020technology}. The video-based learning experience during the COVID-19 pandemic has worsened social exclusion among deaf students, especially with the disruption of daily interactions with other people and inadequate sign language interpreters \cite{aljedaani2022if, tomasuolo2021italian}. Research suggested parents of DHH look for suitable online educational programs, and opportunities for their children to socialize with their DHH peers \cite{kritzer2020educating}. Given current technological advancements, there is a great potential to conduct technology research, targeting inclusion and proper education for DHH students by enabling positive social interactions and collaboration with their peers \cite{batten2014factors}. 



Social media features have proven to facilitate interaction not just among hearing learners \cite{seaman2013social, chen2021afraid}, but for those who are DHH. A survey on predominant social media usage patterns of DHH teenagers in Saudi Arabia revealed that a majority engaged with social media daily \cite{mohammad2023investigating}. Additionally, another survey highlighted how social media empowers DHH students to foster connections and collaborations on information discovery, content creation, and sharing \cite{ali2023investigating}. The positive impact of social media on academic performance was further emphasized by students, who reported heightened interaction, support, and feedback \cite{toofaninejad2017social}. Notably, researchers found that individuals within the DHH community predominantly engage in sharing content through written English, despite a preference for sign language. This inclination can be attributed to challenges such as difficulties in captioning signed videos and obstacles related to recording, such as device limitations, lighting, storage space, etc. \cite{mack2020social,DIS2023}. In summary, these social media features are beneficial for social interaction but have limitations due to the lack of support for the native language of the DHH community - sign language. In our research, we provide social interactions for DHH learners by supporting the use of sign language.

\subsection{Danmaku for Social Interaction and Knowledge Sharing in Video-based Learning} 
Danmaku is a commentary design for online videos that originated in Japan and is now popular on Asian video websites, e.g., Bilibili.com \cite{wu2018danmaku,wu2019danmaku,10.1145/3544548.3581476}. In contrast to traditional video forums, where comments are asynchronously displayed below the video, danmaku comments overlay the video screen. These comments align with specific scenes chosen by online users and scroll from right to left across the video screen. Users can \textbf{anonymously send and view}  danmaku comments in a synchronous way while watching videos. Wu et al. compared danmaku comments with traditional video forums and found the danmaku design significantly promoted user participation and social interactions \cite{wu2019danmaku}. The work also found danmaku and forum complement each other in the realm of knowledge sharing. Danmaku comments primarily facilitate explicit (know-what) knowledge sharing, while forum comments tend to showcase more tacit (know-how) knowledge sharing. He et al. found that viewers were more emotional and expressive in danmaku comments \cite{he2021beyond}. 


Research on online learning using danmaku, found that watching videos with danmaku and providing danmaku is a way for many users to entertain themselves and increase their sense of belonging among other audiences \cite{chen2015understanding,yao2017understanding, liao2023research, lin2018exploratory}. Danmaku allows students to collaborate with their tutors and peers to create learning content by adding comments that help explain learning materials \cite{yao2017understanding}. For teachers, danmaku can bridge the gap between teachers and students, and reduce learners' feelings of loneliness \cite{Li_Zhu_Qian_Ren_Fang_2022}. Conversely, some research pointed out that danmaku is distracting due to its rich abundance of information content and uneven content quality \cite{chen2015understanding}. Furthermore, using danmaku as learner-sourced material raises the possibility that the content created will be ineffective, inappropriate, or incorrect \cite{9353264}.

Meanwhile, research investigating the use of danmaku in video learning with DHH students remains sparse, particularly due to the fact that the current danmaku is text-based which is not fully accessible to DHH people as their native languages are sign and gesture-based languages. Despite accounting for reading proficiency, the differences in background knowledge and information processing strategies make text-based information less accessible for DHH than hearing people\cite{marschark2005access}. Our study addresses this gap by accommodating sign language preferences in redesigning text-based danmaku into signmaku (sign language danmaku). Though signmaku is a new concept, a few known or generally accepted facts support the possibilities for DHH to benefit from synchronous ASL videos along with the original video content. Firstly, many TV shows have a sign language interpreter next to the video content and some video services have a closed ASL interpreting \cite{kushalnagar2017closed, jensema1996closed}. Secondly, instructors believe DHH students have relatively better visual skills than hearing \cite{rodrigues2022unveiling} and DHH students frequently switch their visual attention between the board, slides, instructors, and interpreters \cite{rodrigues2022unveiling}.



\subsection{Anonymization of Sign Language Videos for Privacy Concerns}

The ability to communicate anonymously using one's preferred language holds vital importance across diverse social, professional, and societal settings. As outlined earlier, anonymous comment sharing is a crucial aspect of danmaku interaction and we, too, would need to provide anonymization features in signmaku in our study. While achieving anonymous written communication online is relatively uncomplicated for users of written languages, such options have not been readily available to users of sign languages. For video-based communication, individuals who use spoken languages can conceal their faces on online video-sharing platforms however sign-language users cannot due to the crucial linguistic information conveyed through facial expressions \cite{lee2021american, baker1985facial, coulter1979american, bragg2020exploring}. Respondents expressed willingness to view videos created by individuals with their faces disguised as long as there was an ethical purpose \cite{lee2021american}. 


Sign language is a form of visual communication that employs gestures, facial expressions, and body language. It is a complete first-class language with the same level of complexity as spoken English, but it does not use the same sentence structure \cite{rastgoo2021sign}. Anonymizing sign language videos poses challenges as it's not feasible to directly cover their face. Recent efforts have emerged to assist sign language users in automatically concealing their faces in videos while retaining vital facial expressions; however, this task remains challenging and none of the current approaches offer a sufficiently effective solution. Lee et al.'s work presented several prototypes to DHH ASL signers and revealed users’ perception of key trade-offs among three dimensions-understandability, naturalness of appearance, and degree of anonymity protection \cite{lee2021american}. Clearly representing facial expressions including eyebrow and mouth movements plays a crucial role in conveying content accurately in sign language. A limitation of their study is it did not study sharers' perspective. Additionally, emerging research focused on sign language recognition \cite{radhakrishnan2022cross, rastgoo2021sign}, animation-based generation \cite{saunders2022signing, mehta2020automated}, and machine learning-driven translation systems \cite{tang2021graph, decoster2023}, demonstrating a promising trajectory for the future. To advance the development of sign language processing, the active participation of the Deaf community is imperative across all stages. This involvement is crucial for designing systems that align with user requirements, ensuring usability, and promoting technology adoption \cite{bragg2019sign, prietch2022systematic, articleea}. 

\section{Method}


Inspired by the idea of text-based danmaku~\cite{wu2018danmaku} and its effect of creating a pseudo-synchronous learning experience to engage users \cite{wu2019danmaku}, we introduce \textbf{Signmaku}, which displays peer learners' comments in sign language alongside the video, fostering a pseudo-synchronous co-learning experience. Note that in this study, we used American Sign Language (ASL) as the sign language, and all signmakus were created in ASL.

We conducted a two-phased study to design and evaluate the online learning experience with signmaku. Phase I is a two-round formative design study (N=12) to explore the design parameters of signmaku. Round 1, described in Section \ref{sec:formative} focused on (1) understanding willingness to view (2) filtering styles to maintain anonymity if needed (3) collect ASL data signmakus to be used in round 2. Round 2 focused on fine-tuning the signmaku design. Phase II, detailed in Section \ref{sec:evaluation}, is an evaluation study (N=20) to evaluate how DHH students view and create signmaku in online learning, with a focus on three styles. 

\textit{Positionality of Research Team}
The research team consisted of members with diverse backgrounds, including hearing, hard of hearing, and Deaf individuals: one Deaf research assistant, one hearing research assistant who is a Ph.D. student in linguistics, and an ASL interpreter. 
All study sessions were supervised by a hearing professor with 30 years of experience in teaching DHH students at the collegiate level. The data analysis primarily involved four hearing coauthors who had prior experience in both quantitative and qualitative analysis. They regularly held discussions with Deaf research assistants and the hearing professor to refine the codebook and ensure comprehensive analysis.

\subsection{Phase I: Designing Signmaku--Video Comments in Sign Language} \label{sec:formative}



To delve into the design of video-based learning experience with signmaku, we conducted a university Institutional Review Board (IRB)-approved pilot study with 12 DHH students (8 males, 4 females), including 8 d/Deaf students and 4 hard-of-hearing students, all from a US higher education school for DHH students. Throughout the study, we engaged in interviews and follow-up email conversations with these students to identify the essential design considerations necessary for the concepts to be effective. 

The formative study consisted of two rounds. Round 1 was used to understand the viewing and sharing needs of signmaku and collect data to be presented in round 2. Round 2 was used to iteratively inform the actual design of signamku (Figure \ref{fig:example} shows the final design) via interviews to reveal key design parameters. 
12 participants were recruited in round 1, and several weeks later, round 2 included six of these original participants. In Round 1, each participant was compensated \$40 for their involvement in the 2-hour session. No compensation was offered for follow-up participation in Round 2.  
We analyzed the interview transcript and discussed the findings as well as improvements to the signmaku design at the end of each round. 

\subsubsection{Round 1: Needfinding of Signmaku for Social Connectedness}

In the initial session, the participants were presented with a 15-minute video about Augmented Reality (AR), accompanied by error-free open captions. Following the video viewing, the participants were requested to provide their comments regarding the video content, using either ASL (as signmaku), English, or a combination of both. Additionally, they were asked if they were willing to share their comments for future learners, which would be utilized as supplementary learning material. We inquired about the factors that influenced their willingness to share. Participants were given the opportunity to express their preferences regarding viewing others' comments in either ASL or English as well as the manner in which they would like to access such comments during video learning. On average, each participant made 9 comments, with two-thirds in ASL and one-third in English. ASL-based video comments lasted between 5 and 45 seconds, with longer ASL-based comments including multiple sentences. 

Our round 1 findings indicated that incorporating comments, both as viewers and as providers, has the potential to foster increased social connectedness among students during online sessions by addressing feelings of loneliness that may arise. 
Notably, it was observed that ASL-based comments were generally perceived to be more comprehensible and engendered a stronger sense of connection among DHH participants. Especially as a minority group, DHH community members used their own language to fully express themselves and feel included in the learning process. However, some participants expressed concerns about privacy, because providing ASL-based comments entailed revealing their faces, which can hold grammatical meanings in ASL and cannot be directly removed. Consequently, some participants suggested implementing filters to mask/swap their faces, while others exhibited a preference for more comprehensive masking, such as swapping their entire torso and face. In summary, participants showed a positive inclination towards participating in such commenting activities, and anonymity was essential for some participants.

\subsubsection{Round 2: Specifying Design Features of Signmaku} \label{sec:formative_round2}
Based on the literature review and the results from the first round, we synthesized and learned that signmaku could serve as a valuable paradigm for promoting social connectedness and knowledge sharing. Subsequently, a second round was conducted 1-2 months after the first round to identify key design parameters of signmaku, particularly its display during video consumption. 
For this round, all 12 participants from the previous interview were re-invited and 6 chose to participate. 
We selected 15 signmakus 
from Round 1 and embedded them into the lower right corner of the same 15-minute video about AR. This gave participants on average one signmaku per minute. We then asked our participants to watch this video with signmakus.

The visual layout design underwent several rounds of improvement process based on feedback from each of the six participants. Once participants viewed the video with signmaku, we asked for suggestions to enhance its effectiveness for better comprehension and learning. The draft version is similar to prior literature \cite{chen2022timeline}. In that work, Chen et al. found with the \textit{embedded} layout, learners altered their attention more frequently between two comments and remembered more surface-level comments. Therefore, we selected the \textit{embedded} layout for signmaku.

The participants' feedback was carefully incorporated into the design, and subsequent emails were sent to them to confirm their willingness to provide additional inputs. The original layout for video-watching underwent multiple refinements to arrive at the final layout used in the subsequent formal study. The final design of signmaku is shown in Figure \ref{fig:example}. Below, we provide an outline of the key parameters that were manually adjusted during the improvement process, along with the reasons behind those adjustments. Round 2 was dedicated to adjusting the parameters for our second study; however, these parameters might not be universally applicable to signmaku in different video contexts. Future research should explore methods for automatically completing the adjustment of design parameters. 

\begin{itemize}
    \item \textbf{Duration}: 
    Participants were required to alternate between reading signmaku and captions while also attending to the video's visual content. Consequently, it was crucial for the signmaku duration to strike a balance. If it was too lengthy, it would interrupt the participants' \textbf{comprehension flow and attention span}, but if it was too short and limited to a single sign, it might ineffectively convey the intended message. Thus, we reduced recommended duration from 30 to 10 seconds. This allowed DHH individuals to complete signing a full sentence in ASL. It is important to note that this parameter was tailored and refined based on the pace of the selected video content. It may require further exploration and adaptation when applied to different video contexts with varying tempos.

   \item \textbf{Dimension}: Participants found it crucial for the signmaku to be sufficiently large, ensuring clear visibility of the signer's \textbf{facial expressions, fingers, and elbows} for the participants to understand the language. Simultaneously, it is essential to avoid overlap with captions and video content. To address this, we positioned the signmaku in the lower right corner, aligning it with the height of the text captions. This placement allowed learners to seamlessly \textbf{switch their visual focus} between the two sources with ease. In the initial setup, an additional panel for text-based comments was incorporated alongside the video panel. This panel automatically scrolled during video playback while showing text-based peer comments. However, we later removed this panel from the formal study to provide a larger video player that could ensure better visibility of signmaku and captions.

   \item \textbf{Placement in relation to the video content}: Participants expressed a desire for the signmakus to accurately correspond to the video's content, particularly when referencing specific elements like equations. However, they also suggested that the pop-up does not necessarily need to appear directly next to the referenced content. They felt that the unpredictability of where the signmaku might appear could add to the confusion and pressure, leading them to prefer a \textbf{consistent presentation}, allowing them to know clearly where to expect the signmaku. 

  \item \textbf{Categories}
  : We found that some signmakus from round 1 focused more towards complaints of video content and less towards the knowledge of the video (e.g., ``When can this video end, I am tired.''). For knowledge-sharing purposes, we manually selected signmakus that commented on the actual video content instead of emotion-only content, in a similar approach to previous danmaku research \cite{wu2018danmaku} (e.g., ``The prototype would not fly nowadays. But I will admit it does look pretty cool.'')
   
  \item \textbf{Two Filters vs Unfiltered}: 
   Following participants' feedback, we applied filters on signmakus with advanced face-swap technology, similar to a prior study \cite{lee2021american} that used several prototypes to automatically disguise the face in ASL videos while preserving essential facial expressions and natural human appearance.
  We explored multiple technologies of face filtering and eventually decided to examine in-depth the two most promising filters. 
  The first filter was the \textit{cartoon} filter using VToonify \cite{yang2022vtoonify}, a portrait style transfer tool, which retains facial movements while giving the face a cartoon-like appearance and providing a basic level of privacy. The original human hands are preserved in this filter. The second filter was the \textit{robotic} filter using DeepMotion, an AI-based motion capture and body tracking tool, which masks the entire body and utilizes artificial hands and faces to create a robotic appearance. This filter provides an increased level of privacy than the \textit{cartoon} filter. 

    \item \textbf{Quantity}: 
    Upon further discussion, we retained 15 signmakus with the average of one signmaku per minute. With 2 filter options and unfiltered signmakus, this allowed us to split the signmakus into 3 groups of 5 signmakus and distribute them evenly across three 5-minute segments of the video, i.e., the beginning, the middle segment, and the final minutes of the video. For each group of signmakus, we included diverse emotions  (two negatives, two positives, one neural).
    Notably, participants mentioned how having all 3 styles of signmakus randomly popping up could be distracting as the viewer would never know which style would appear next. Therefore, each group had the same style before switching to the next style of signmaku, so that participants would only encounter two changes in styles of signmakus throughout the video.

    \item \textbf{Quality and Clarity}: The signing quality and clarity were manually assessed by four researchers on the team. When creating filtered signmakus from round 1, we noticed variations in video signing quality and clarity among participants. Some participants' filtered signmakus were difficult to understand.
    Upon further discussion, we retained signmakus from only one of the participants that had best quality and clarity in both unfiltered and filtered signmakus. This participant had made 11 signmakus that spread across the 15-minute video timeline, and we cut their longer ASL recordings into several signmakus and remapped smaller signmakus back to the video timestamp based on the contents in her signmakus. 
    Other factors related to signing quality and clarity were left for future exploration. 
   
\end{itemize}

\begin{table*}[h!]
    \centering
    \caption{Demographics of the participants. All participants were recruited from a university that focuses on education for DHH. Each participant experienced the three conditions, featuring stylistically relevant styled signmaku comments, in a randomly assigned order while watching the video.
    }
    \begin{tabular}{cccccc}
    \hline
    \textbf{PID} & \textbf{Gender} & \textbf{Hearing Status} & \textbf{Age} & \textbf{Ethnicity} & \textbf{Randomly Assigned Conditions} \\ \hline
        \textbf{T1} & Male & Deaf & 32 & Black & Cartoon - Robot - Raw \\ \hline
        \textbf{T2} & Male & Hard of Hearing & 19 & Black & Raw - Robot - Cartoon \\ \hline
        \textbf{T3} & Female & Deaf & 19 & Asian & Cartoon - Robot - Raw \\ \hline
        \textbf{T4} & Female & Deaf & 22 & White & Raw - Robot - Cartoon \\ \hline
        \textbf{T5} & Female & Deaf & 43 & White & Raw - Robot - Cartoon \\ \hline
        \textbf{T6} & Male & Deaf & 22 & Portugal  & Robot - Raw - Cartoon \\ \hline
        \textbf{T7} & Male & Deaf & 19 & White & Robot - Raw - Cartoon \\ \hline
        \textbf{T8} & Male & Deaf & 22 & Multiracial & Robot - Raw - Cartoon \\ \hline
        \textbf{T9} & Male & Deaf & 22 & White & Robot - Cartoon - Raw \\ \hline
        \textbf{T10} & Female & Deaf & 21 & Multiracial & Cartoon - Raw - Robot \\ \hline
        \textbf{T11} & Female & Deaf & 31 & White & Cartoon - Raw - Robot \\ \hline
        \textbf{T12} & Male & Deaf & 18 & White & Robot - Cartoon - Raw \\ \hline
        \textbf{T13} & Male & Deaf & 28 & White & Cartoon - Raw - Robot \\ \hline
        \textbf{T14} & Female & Deaf & 20 & Multiracial & Raw - Cartoon - Robot \\ \hline
        \textbf{T15} & Male & Deaf & 19 & White & Robot - Cartoon - Raw \\ \hline
        \textbf{T16} & Female & Deaf & 43 & White & Raw - Cartoon - Robot \\ \hline
        \textbf{T17} & Non-binary & Deaf & 28 & Multiracial & Raw - Cartoon - Robot \\ \hline
        \textbf{T18} & Male & Deaf & 22 & Multiracial & Raw - Robot - Cartoon \\ \hline
        \textbf{T19} & Male & Hard of Hearing & 26 & White & Robot - Raw - Cartoon \\ \hline
        \textbf{T20} & Male & Hard of Hearing & 31 & White & Cartoon - Robot - Raw \\ \hline
    \end{tabular}
    
    \label{tbl:demographic sheet}
\end{table*}

\subsection{Phase II: Comparing Three Styled Signmaku} \label{sec:evaluation}
Following the formative study, we conducted experiments to evaluate the impact of differently styled signmaku on users' experience during video-watching (RQ1) and creating new signmakus after video-watching (RQ2). As mentioned in section \ref{sec:formative_round2}, we chose two filter styles to address privacy concerns when sharing signmakus with other learners: \textit{robotic} and \textit{cartoon}. We provided a third option, \textit{realistic}, to serve as a privacy-agnostic comparison without applying any filters to the signmaku.


\subsubsection{Participants Information} 

A total of 20 DHH participants were recruited. All participants in our study were students currently enrolled in college, ranging in age from 18 to 43. All participants were from a private university in the US focusing on education for DHH students. 
None of the participants were involved in the formative study described in Section \ref{sec:formative}. Out of the 20 participants, seven were identified as female, one as non-binary, and the remaining 12 as male. The majority, 17, identified their hearing status as d/Deaf within the DHH community, while the other three participants identified as hard of hearing. Seven participants were from a family where no immediate family member were DHH, whereas the other 13 participants had one or more DHH family members. All participants had learned ASL for at least one year, including 11 participants who started learning ASL from birth. 
Regarding their classroom language preferences, 11 participants felt comfortable with using ASL only, one participant felt comfortable with English only, and the other 8 participants were comfortable with both ASL and English in the classroom. As for ethnicity, 2 participants identified as Black, 1 as Asian, 11 as White, 5 as Multiracial, and 1 as Portuguese. Most were non-STEM majors e.g., business. See Table \ref{tbl:demographic sheet} for the demographics of the participants. Each participant was compensated at 20 USD per hour. The duration of each study ranged from one hour to two hours. The study plan was approved by the university IRB.


\subsubsection{Experimental Design}

\begin{figure*}[t]
    \includegraphics[width=.8\textwidth]{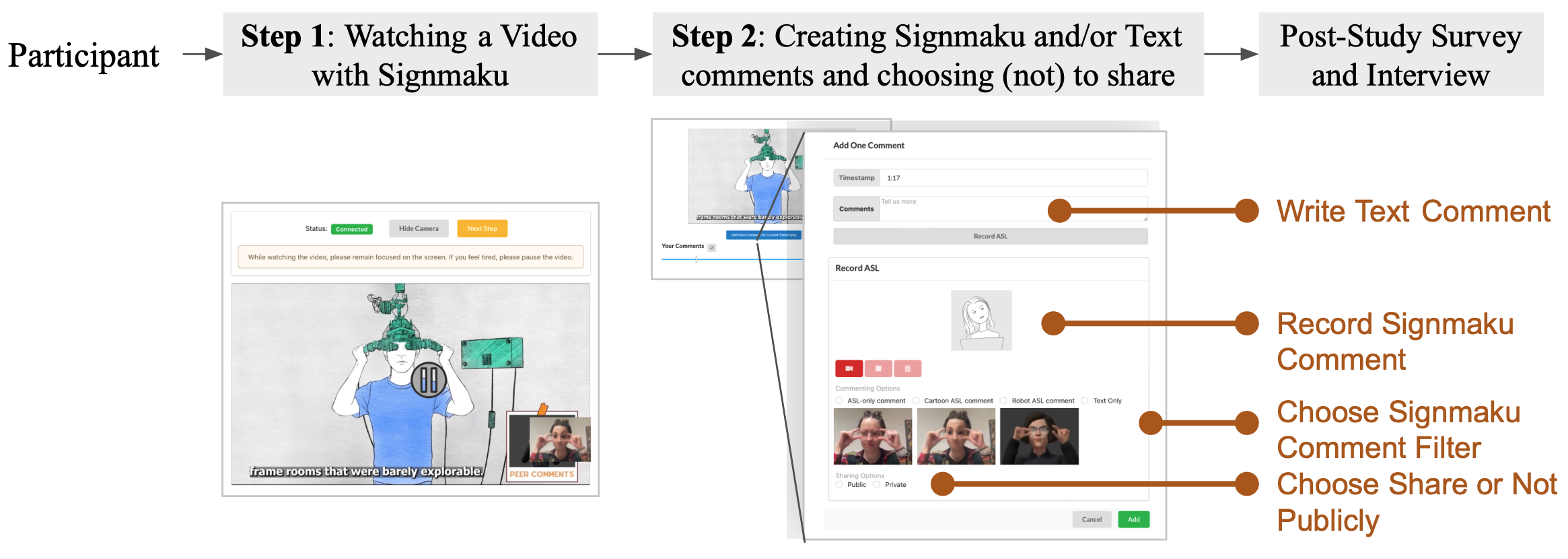}
    \caption{
    Experiment Design. Each participant completed two activities 
    during the study. 
    First, they watched a video about augmented reality (AR) with error-free open captions and signmakus (RQ1). Second, they provided their comments in signmaku (ASL comment) or text comments (RQ2). A comment may be consisted entirely of text, solely use ASL, or, on very rare occasions,  include a mix of both. They then completed a post-study survey and interviews. Participants were ask to provide signmaku only after they finished watching the video for the first time.}
    \label{fig:Processfig}
    \Description{This image demonstrates the two activities of our user study processes: step 1 - watching a video with signmaku, step 2 - creating signmaku and/or text comments and choosing whether or not to share them. The study was followed by post-study surveys and interviews. There are also two screenshots of the interface for each activity. For step 1, the interface consists of control of automated emotion recognition as well as the video with open caption and signmakus. For step 2, the diagram shows the "Add One Comment" interface after pressing a blue button. From top to bottom, this interface allows users to write text comments, record signmaku comments, choose signmaku comment filters, and choose to share or not publicly. The comment filter consists of four options: realistic, cartoon, robotic, and text comments. At the bottom right of the interface, there is a green button where the user can save the comment.}
  \end{figure*}

We conducted a two-step user study, as shown in Fig. \ref{fig:Processfig}. The first step addressed RQ1, while the second step was to address RQ2. The study was followed by a survey and concluded with an interview. The prototype used for this study was developed based on the proposed design from the formative study, as shown in Fig. \ref{fig:Processfig} (outcome of round 2).
To help participants navigate through the prototype and study procedure, we prepared instructions in a pre-recorded ASL video, spoken and written English to the participants throughout the study. 
Participants could also ask questions in ASL or by typing in the chat box in Zoom. The researchers turned off their cameras while participants completed each step to avoid distraction and free up visual space, but cameras were turned on if the participant requested help or strayed from the procedure. All but one participant preferred to use ASL, while the remaining participants preferred to use spoken English. 
During the onboarding process, participants were introduced to the overall goals, procedural steps, and visual format, (e.g., where and how signmaku would pop up). This preview gave participants a clear idea of where to look and what to expect. Each step of the procedure was reintroduced before each section, granting the participants additional opportunities to review the process and ask clarifying questions. The Deaf individual and hearing research assistants with interpreting experience took turns leading 19 out of 20 user study sessions (T1 - T19), and all recordings were transcribed by these same assistants. The remaining session was led by the first author with an interpreter. 

\textbf{Step 1 - Video Watching with Signmaku.} First, participants watched a 15-minute error-free open-captioned 
video about AR, the same video used during the formative study described in Section \ref{sec:formative}. This step allowed us to observe how participants viewed the signmakus. Participants kept their webcams turned on and consented for AI to recognize their facial expressions while watching the video. However, participants could opt to hide their webcam viewer on their interface if they did not wish to see themselves while watching the video. 
During the video, 15 peers' signmakus, as collected from the formative study, appeared on the bottom right of the video - on average one signmaku per minute similar to the formative study. Each signmaku popped up in real-time and related to the information in the video at that moment. 

During the formative study, participants suggested using filters to preserve privacy while keeping the content in a signmaku. To explore how the styles of signmakus may impact how participants perceive these signmakus, we used the \textit{cartoon} and \textit{robotic} filters, in addition to the \textit{realistic} unfiltered signmakus, as shown in Fig. \ref{fig:Processfig}. There were 15 total signmakus, grouped into three sets of five signmakus, that each participant viewed at the same timestamps in the video. We detailed the process of choosing and refining signmakus used in this phase of the study in section \ref{sec:formative}. The order of appearance of the styles in the three sets of signmakus was randomized. Table \ref{tbl:demographic sheet} includes the order of styles received by each participant. 

\textbf{Step 2 - Creating Signmaku during Video Reviewing.} After finishing the video for the first time, participants could review the video in part or in whole in order to add their own signmaku or text comments. This step provided insights into what options participants might consider when sharing signmaku with other learners. Participants saw all three signmaku styles before they decided which style they wanted to share. This provided an informed selection and separated the viewing (RQ1) and sharing (RQ2) effects of signmaku individually. Participants were allowed to add text comments, which emphasized the differences between recording an ASL comment and typing an English comment. 

The video with embedded signmakus was identical in both step 1 and 2. When adding a comment, participants could jump directly to the desired timestamp and select a language preference: type a written English comment or record an ASL signmaku comment. We encouraged participants to add at least one signmaku and one text comment, but they were not required to add both. Signmaku selections had two additional options: filters and sharing. Participants could apply one of three filters: \textit{robotic}, \textit{cartoon}, and \textit{realistic}. Signmakus could be made public or kept private. Such customization enhanced participants' privacy control over each comment, which was a concern raised in the formative study.

Note that the focus of our study is on understanding user behavior and does not aim to make any technical contribution. Our prototype did not implement real-time filtering due to the lengthy processing time. If time permitted during the interview, some participants' ASL comments were processed outside our current prototype, allowing them to provide additional feedback on the filters.


\subsubsection{Measurements and Data Analysis} For each user study conducted to gather and assess feedback on the prototype, we collected the following data: responses from a post-study survey, emotions detected via webcam during video viewing, action logs, and interview responses.  

\textbf{Post-Study Survey:} The survey consisted of two major parts: Part 1 focused on the viewing experience (RQ1), and Part 2 concentrated on the creating experience (RQ2). Additionally, we collected participants' demographic information as well as their preferred language in learning. Please refer to the appendix (Appendix \ref{appendix:survey}) for full survey questions. 

\textit{Part 1- viewing experience:} Participants were asked to rate their subjective task load \textit{while} viewing the three signmaku styles. We focused on participants' opinions toward the task load (mental demand, physical demand, temporal pressure) \textit{while} watching videos with different styles of  signmaku. We adapted NASA Task Load Index \cite{hart1988development} (Likert scale 1-10), as it has been used to study DHH interactions with technology, e.g., \cite{li2022soundvizvr,dust2023understanding}. Each participant responded to nine questions, divided equally among three styles (three questions per style). For example, a sample question was ``How mentally demanding was viewing the pop-up ASL-only comments?'' 

After collecting the survey scores, we performed three repeated measures ANOVA tests with a randomized effect to compare the effect of signmaku style on (1) mental demand, (2) physical demand, and (3) time pressure. A random effect, ``1/PID,'' was included to account for individual differences among participants that could not be explained by the fixed effects in the model \cite{chang2023citesee}, where ``PID'' was Participant ID as shown in Table \ref{tbl:demographic sheet} column 1. The power analysis was conducted in R using the WebPower package. Note that all statistically significant tests with this part of survey data remained significant (p\textunderscore adj < 0.05) after a Bonferroni correction; we still used ``p'' instead of ``p\textunderscore adj'' when reporting data analysis in the findings.  

\textit{Part 2 - creating experience:} Participants were asked to answer 20 questions about their perceived sense of community when using our prototype design to view and share to answer RQ2. These questions were based on the Classroom Community Scale \cite{rovai2002development, dawson2006study}. After collecting participants' ratings, we first added up the survey responses to two composite scores of sense of community in learning - ``sense of connectedness'' and ``sense of learning''. Then, we ran correlations between survey responses on task load and survey responses on sense of community, because we were curious if the increased load of signmaku, especially \textit{robotic} found in RQ1 (presented in section \ref{nasa}), had any positive or negative relation with the overall experience. We first ran the correlation between two composite scores and each style's three questions for task load. The results were overall similar across the three styles, so we further conducted correlation analyses on each of the 20 questions. We compared each of the 20 questions with the three task load questions specific to each style.

\textbf{Learners' Emotions Automatically Detected: } 
Prior works showed that learners' emotions could be estimated using facial recognition technologies continuously \cite{noroozi2020multimodal, lasri2019facial}. In our study, we not only gathered data on participants' subjective perceptions but also utilized facial recognition technology, as described by \cite{deng2020multitask}, to identify their emotional responses during the learning process. To address RQ1, we conducted a statistical analysis to examine changes in participants' emotions before and after they were exposed to various styles of signmaku in Phase 1. Specifically, we looked at the ``valence changes'' in participants' emotional valence triggered by the appearance of a signmaku on the video display. We calculated the valence change by taking the value from the first two seconds of the signmaku popup and subtracting the average valence value over the entire 15 minutes of video viewing. 

Each participant contributed a total of 15 emotion expression changes, comprising five times for each of the three styles. We utilized repeated measures ANOVAs to assess the impact of signmaku style on valence changes, which ranged from -1 to 1 (representing a spectrum from negative to positive, with 0 indicating a neutral state). To account for individual differences among participants, ``1/PID'' was also included as a random effect in the model, similar to the survey analysis. Since these technologies are not always 100\% accurate in real-life applications, we provided participants with visualizations of their emotions over the video timeline, derived from facial data (as in \cite{DIS2023}). At the beginning of the interview, we explained the data usage, allowing participants to review it and opt-out if desired. Participants were generally comfortable with our approach, suggesting it was adequate to conduct within-subject comparison.  Note that all statistically significant tests with facial data remained significant (p\textunderscore adj < 0.05) after a Bonferroni correction; we still used ``p'' instead of ``p\textunderscore adj'' when reporting data analysis in the findings.

\textbf{Action Log:} We used participants' action logs to understand the learning behaviors when providing signmaku. More specifically, we collected the following logs: the written text in English, the duration of typing text, the duration of crafting each signmaku, the number of re-recorded signmaku, the option of choosing the three styles of signmaku filter, and the choice of sharing or not.  
We compared counts of comments between the three types using Wilcoxon Signed-Rank Test, with the results presented in section \ref{1}. We also compared the time spent per person and per clip on private/shared options using the same text, with the results presented in section \ref{2}. Due to our limited sample size, we did not conduct a further statistical comparison of effort for each type in the context of private vs public options. Additionally, we compared time spent per person and clip on ASL/typed English options using the same text. The details of this comparison are presented in section \ref{3}.

\textbf{Post Study Interview:} During the semi-structured interviews, participants were asked about their thoughts when using the prototype, their reasoning behind certain actions during previous steps, and their suggestions toward prototype improvement. Participants either opted to use ASL or spoken English to complete the interview. Sample interview questions included, ``\textit{which filter do you prefer or not prefer when viewing others’ signmakus and why};'' ``\textit{when sharing your signmakus to others, do you have any concerns about two filtered options}.'' 
A thematic analysis of the transcriptions extracted the main themes \cite{thomas2006general}. Two coauthors independently coded three of the transcripts and discussed together to create an initial version of the codebook. One coauthor further completed the remaining transcripts. The two coauthors discussed frequently to revise the codebook and group codes into themes. Some sample codes include positives and difficulties of viewing or sharing signmakus, strategies adapted when using the interface, and accessibility or usability suggestions. 





\section{Findings} \label{sec:finding}


\subsection{Engaging Learners through Viewing Three Styles of Signmaku (RQ1)}  
Overall, the majority of the interviewed participants reported that signmakus were able to reengage and amuse them after being tired, because reading text-only captions could be \textit{``tedious''}, \textit{``boring''} and \textit{``tiring''}. And the signmakus made participants feel more connected, e.g., they wondered \textit{``what's the next comment would say''}, and felt they \textit{``we were watching the video together at the same time...not just passively watching''} (T1). The signmaku provided participants more visual information via ASL when feeling \textit{`'struggled...like science, with big words,''}, which in turn increased their motivation: \textit{``the peer comment actually helped with “visualizing”... ASL helps me learn more!''} (T15). T11 hoped to have signmaku for long videos to make the learning process more engaging and motivating. 

When asked if they found competition between different information sources, some disagreed and further said they used different data sources to double confirm their understanding and to ``\text{fill in the blank.}'' T11 specifically mentioned, as DHH they are used to and is good at using different information sources to guess and disambiguate. The signmaku is an additional information source that provided cues to support the process. Notably, some DHH learners shared that they developed strategies to pay attention to multiple visual elements in videos, allowing them to compensate for the lack of auditory information.  

However, challenges remained when watching videos with signmaku. Interview data revealed that the main challenges agreed upon among participants were to quickly form their own new visual attention management strategy to fully comprehend the video content together with captions and signmaku. They would constantly switch visual attention to viewing video content and caption and signmaku. They considered the ability to understand both at the same time as a skill that needed practicing: \textit{``I can see the video and peers’ comments simultaneously, but I am not that skilled. It is challenging and distracting... Need practicing''} (T16). T4 also described a conflict in attention between caption and signmaku, \textit{``It was interesting when I saw the ASL comments. I was excited to see and watch those and I just stopped looking at the captioning. Yeah... When I was reading the comments, I found myself agreeing...''} 

We next present a detailed comparison between the three styles of signmaku.

  \begin{figure}[!t]
    \includegraphics[width=.45\textwidth]{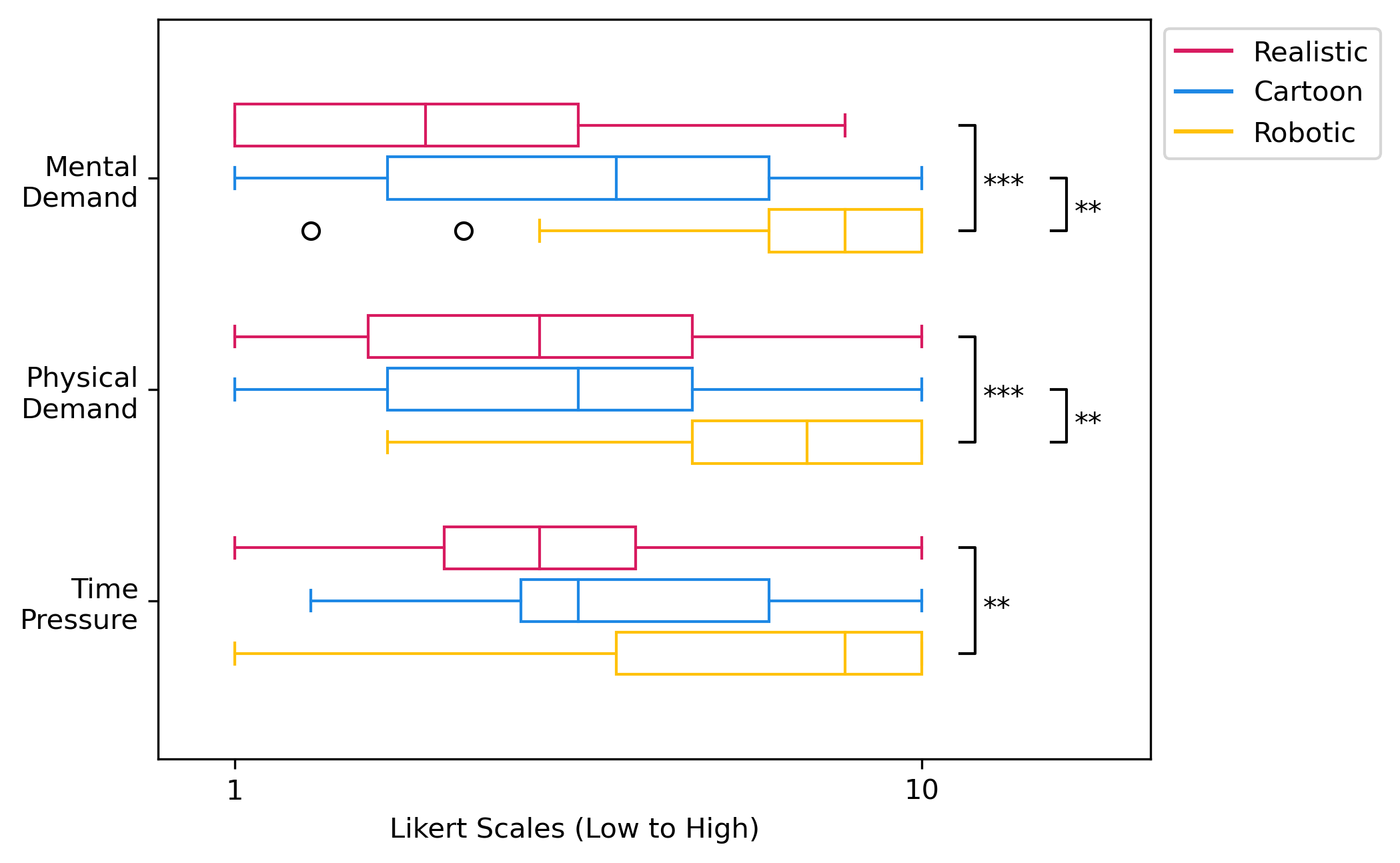}
    \caption{Boxplots of participants' reported mental demand, physical demand, and time pressure of viewing three styles of signmakus (N=20) on 10-point Likert scales (Low to High). 
    The mental demand, physical demand, and time pressure were all significantly higher when viewing \textit{robotic} ASL signmakus compared to the realistic styles. Note **, *** signify p< .01 and .001, respectively. 
    } 
    \label{fig:boxplot}
    \Description{There are three groups of boxplots in this figure. Each group represents the survey measurements of mental demand, physical demand, and time pressure. Each group has three boxplots representing the three ASL filters (realistic, cartoon, and robotic). For mental demand, robotic signmakus are significantly higher than realistic signmakus (p< .001) and cartoon signmakus (p< .01). For physical demand, robotic signmakus are significantly higher than realistic signmakus (p< .001) and cartoon signmakus (p< .01). There is no significant difference between realistic and cartoon signmakus in physical demand. For time pressure, realistic signmakus are significantly lower than robotic signmakus (p< .01). Pairwise analysis tests did not reveal any additional statistically significant differences between realistic and cartoon signmakus, nor between cartoon and robotic signmakus.}
  \end{figure}

\subsubsection{Significant differences in perceptions of viewing the three signmaku styles} \label{nasa} 

Participants were asked to rank the three styles of signmaku for 'enjoying the signmaku with the video.' The majority (66\%) selected \textit{realistic} as their first choice, 33\% chose \textit{cartoon} as their first choice, and 90\% considered the \textit{robotic} version as the least enjoyable. To investigate further, we conducted a statistical test to determine if the different styles of signmaku had different task demands when viewing. Three repeated measures ANOVA tests with a randomized effect were performed to compare the effect of signmaku style on (1) mental demand, (2) physical demand, and (3) time pressure. A power analysis found sufficient statistical power for the study (0.81).  Results are shown in Figure \ref{fig:boxplot}.

\textbf{Mental Demand}: There was a statistically significant difference in mental demand (F(2,57)= 17.25, p< .001) between at least two groups. On average, participants found the mental demand of viewing \textit{realistic} to be very low on 10-point Likert scales (M=3.5, SD=2.7). The mental demand for the \textit{cartoon}, was slightly higher (M=5.7, SD=2.9) than \textit{cartoon}. The \textit{robotic} had very high mental demand (M=8.4, SD=2.3). A post hoc analysis found that the \textit{robotic} signmakus had significantly higher mental demand than the \textit{cartoon} and \textit{realistic} signmakus (p< .01 and p< .001 respectively). The \textit{cartoon} had similar mental demand as \textit{realistic}. 

\textbf{Physical Demand}:  There was a statistically significant difference in physical demand (F(2,57)= 10.32, p< .001) between at least two groups. For the \textit{cartoon}, the physical demands (M=5.3, SD=2.6)  were medium and similar to \textit{realistic} (M=4.7, SD=2.7). The \textit{robotic} had very high physical demand (M=8, SD=2.1). A post hoc analysis found that the \textit{robotic} signmakus had significantly higher physical demand compared to \textit{cartoon} and \textit{realistic} signmakus (p< .01 and p< .001 respectively) .

\textbf{Temporal Pressure}:  There was a statistically significant difference in temporal demand (F(2,57)= 5.983, p< .01) between at least two groups.  For the \textit{cartoon}, the time pressure (M=6.0, SD=2.6) of viewing was similar to \textit{realistic} (M=4.9, SD=2.6), and participants felt neither rushed nor relaxed. The \textit{robotic} made participants feel very rushed (M=7.8, SD=2.8) had higher temporal pressure than \textit{realistic} (p< .01) (post hoc analysis).


\subsubsection{Realistic ASL yielded the least cognitive load, followed by Cartoon ASL} \label{understand}


 
Interview findings explain that real hands in both \textit{cartoon} and \textit{realistic} signmaku greatly contributed to the understandability of the content, which was absent in \textit{robotic}. T1 explained that hands and lips heavily need improvement for \textit{robotic}. T4 found \textit{cartoon} and \textit{realistic} both looked like real human signing, but they ``\textit{didn’t notice that it was at all human.}'' and thought ``\textit{the robot is very hard to understand, hands are completely off...}''


Additionally, the use of facial expressions was key to helping one's understanding of the signed content including emotions, which was suppressed in \textit{cartoon} compared to \textit{realistic}. While the limitations of facial expressions of \textit{cartoon} were noted by some of the participants. For example, T10 found  \textit{cartoon} understandable and commented on some facial movements: 

\begin{quote}
    \textit{``I liked the cartoon. The (cartoon's) signing was clear but it really wasn’t the same as the real ASL signing.  It’s harder because the clarity of the cartoon’s facial expressions was harder to understand. It (cartoon) wasn’t as clear. But the (realistic) ASL comments were clear, and those facial expressions were clear.''}" 
\end{quote}

When experiencing with 
\textit{robotic} signmaku, participants created the metric, ``smoothness,'' for measuring how well signing avatars presented different modalities together. 
T15 commented: ``\textit{The robot was not natural. It needs more flow. Not choppy. I wasn't sure what it said.}''  T3 also complained: \textit{``The robot is stiff, not flexible like a real person.''} T16 provided a more detailed explanation: 

\begin{quote}
    \textit{``I don’t like a robot avatar at all. It is useless because it is chunky, and it is hard to understand its signing. I could not recommend using it at all for ASL. I learned an interesting tidbit from Dr. Annelies  Kusters, who mentioned how you perceive sign language in texture. That got me thinking about ASL’s texture, and it got me to understand that ASL is very smooth, flexible like water. So the avatar for ASL should be curvy and smooth, whereas the robot is the opposite.''}
\end{quote}
  
\subsubsection{Inaccurate \textit{Robotic} ASL triggered the highest cognitive load, compared to \textit{Cartoon} and \textit{Realistic} being similar} 
Overall, \textit{robotic} was hard to understand with fake and inaccurate hands and faces, which inclined some learners to zoom in and enlarge the signmaku--trying to see the signs more clearly.  Also, the hands were using time and space awkwardly, making it hard to interpret semantic meaning. This was also associated with missing the chance to look at the original video content and the sense of pressure associated with feeling left out. Some eventually decide to ignore the \textit{robotic}, as T5 explained: 

\begin{quote}
    \textit{``with the robot, I really had to concentrate. I didn’t understand it at all. It wasn’t worth the work. I felt like it ate up all of my energy that should have been on the content video. So, I started completely ignoring the robot. I continued to monitor the comments, but if it was the robot, I wouldn’t look. I’d continue to look at the video.''}
\end{quote}

Some agreed viewing \textit{robotic} required too much cognitive load that they intentionally decided to ignore it and to prioritize focus on the captions, which were more commonly mentioned for \textit{robotic} than the two other styles. Some further suggested giving DHH learners more control so they can better know what to expect and where to look as they said without control \textit{``I will be lost and I don’t know what to look at''} (T16). Some participants were attracted by the signmaku and forgot to look back at the captions. T4 found \textit{robotic} signmaku to be challenging to understand and eventually ignored it:

\subsubsection{\textit{Cartoon} signmaku brought a lot of joy.}

While participants watched the video, their facial movements were collected to estimate their emotional responses~\cite{deng2020multitask}. There was a statistically significant difference in valence (negative to positive) changes between at least two signmaku styles (F(2,247)= 4.078, p< .05 ), indicating that the emotional changes from negative to positive was influenced by the style of signmaku watched. No significant difference was found for arousal changes between signmaku styles. The power analysis indicates a good statistical power for the study (0.84).  

Post hoc analyses revealed that \textit{cartoon} increased positive emotions (M= 0.16, SD= 0.08 ) more than \textit{robotic} (p< .05) (M= 0.008, SD= 0.05) and \textit{realistic} (p< .05) (M= 0.007, SD= 0.05); and no significant differences were found between \textit{robotic} and \textit{realistic}. In summary, these results suggested that watching \textit{cartoon} signmakus brought a positive emotional effect compared to watching \textit{robotic} or \textit{realistic} signmakus. 

This result was aligned with participants' interview feedback, that the most common reason for \textit{cartoon} to be preferred was that it felt fun to watch, as it \textit{``reminds me of Disney due to big eyes.''} (T11). \textit{Cartoon} was also regarded with high potential for customization and a wide range of choices it could offer, such as creating funny or role-playing content in varied learning and social contexts. For instance, T4 commented: 

\begin{quote}
``\textit{...sometimes you can become an animal. Or your head can become a character? Or it’ll make your face and body a different color...funny.}'' 
\end{quote}

T16 also added that cartoon signmakus might not be preferred if the video was for a serious context, e.g., a meeting or business: 

\begin{quote}
    \textit{``The cartoon-style comments are fun to use. It depends on the seriousness of the context. Keeping a cartoon avatar would be a good idea to make the learning experience fun learning. For children, playing is learning, and it applies to any age. If you enjoy and you will enjoy learning. Using cartoon avatar depend on the context. In an informal environment, it could be fun to use.''}
\end{quote}

\textbf{Summary-RQ1} 
While signmaku, video, and captions might compete for DHH's visual attention, they provide complementary and visual information for disambiguating captions that are boring and tedious to follow. Participants' survey results showed that viewing \textit{robotic} put the highest task load (mental demand, physical demand, time pressure)  on them. According to their interview feedback, the \textit{robotic} signmakus were the least accurate in delivering hand movements and facial expressions. On the contrary, \textit{cartoon} signmakus did not pose more physical demand and time pressure than the \textit{realistic} ones. Moreover, \textit{cartoon} signmakus were favored by participants for their entertaining effect, which aligned with participants' facial expression changes captured by the automatic emotion recognition tool. 


\subsection{Engaging Learners via Creating and Sharing Signmaku (RQ2)} RQ2 reveals users' perceptions of creating and sharing signmaku \textit{after} viewing signmaku in RQ1, using survey, system log analysis, and interview. We compared the user experience of creating and sharing ASL comments to text comments, uncovering design opportunities. In total, our participants created 68 signmakus with an average of 3.4 signmakus per participant (SD= 3.8). 


\subsubsection{Creating and Sharing signmaku promoted a sense of learning community} During the post-study survey, participants reported an increase in ``sense of connectedness'' (average score= 33.8/50, SD= 3.9) and ``sense of learning'' (average score= 28.8/50, SD= 5.9), when they were asked to compare the proposed design against video-based learning without viewing or creating signmaku themselves.  
Among all three styles, increased mental and physical demand positively correlates with a ``sense of connectedness'' but not with temporal pressure. 
Increased physical demand positively correlates with ``sense of connectedness'' among \textit{realistic} (\textit{$\rho$}= 0.61, p< .01) and \textit{cartoon} (\textit{$\rho$}= 0.50, p< .05), but not for \textit{robotic} (\textit{$\rho$}= 0.25, p> .05). Increased mental demand positively correlates with ``learning opportunities'' among \textit{realistic} (\textit{$\rho$}= 0.50, p< .05) and \textit{cartoon} (\textit{$\rho$}= 0.53, p< .05), but not for \textit{robotic} (\textit{$\rho$}= 0.33, p> .05). Further research should understand why there appear to be differences among the three styles.
 


The interview findings suggested that being able to create and share comments also fostered a feeling of connectedness. For example, T1 was one of the many participants who re-watched all signmaku once again when providing their own comments and tried to reply to others. 
When asked why, T1 said, 

\begin{quote}
    \textit{``...I felt connected. It was like, `I see you!' and I wanted to add to that in my comment. I felt like I could relate. `I feel you...' Connected to that comment... Enough that you felt motivated to provide your own comment...''} 
\end{quote}

\subsubsection{\textit{Cartoon} offered a balance between anonymity and understandability.} \label{1}
When creating signmaku, the options for \textit{realistic} and \textit{cartoon} were preferred over \textit{robotic}. 
User interaction action logs recorded that our participants (N=20) created 33 \textit{realistic} signmakus, 25 \textit{cartoon} signmakus, and 10 \textit{robotic} signmakus. There were significantly fewer \textit{robotic} comments than the two other styles (p< .05). Among the three styles, 25 uploaded \textit{cartoon} comments of which 11 were private and 14 public, while for \textit{realistic} 21 were private and 12 were public, and for \textit{robotic} 3 were private and 7 were public, respectively. In each style, approximately half of the comments were selected for sharing, and half were kept private. Regarding the sharing rate of various signmaku styles, there were no significant differences among the three styles when using a chi-square test. The limited sample size did not support further comparative analysis of each style. 


The interview results found that the \textit{robotic} style prioritized privacy at the expense of understandability in sharing, while the \textit{cartoon} style was an appropriate compromise between both requirements. In contrast, the \textit{realistic} style failed to maintain any level of privacy.  Better ASL understandability is necessary for self (used as person note-taking for further reviews) and others (sharing publicly). For example, T15, who found \textit{robotic} signmakus more mentally challenging and harder to interpret than \textit{cartoon} and \textit{realistic}, switched 
between the two styles but eventually decided to use \textit{robotic} for anonymity when providing signmakus: 
\begin{quote}
    ``\textit{... I wasn't sure what it said, it's a lot of work...(for sharing) I prefer that one [realistic] because feels more natural, like peer-to-peer...However, for those people who don't feel comfortable revealing themselves, I would recommend the robot.. I think the robot is the best (for staying anonymous in sharing).}''  
\end{quote}

While having good understandability, \textit{cartoon} also preserved an acceptable anonymity ability for some participants. T3, T5 T7, and T11 only made \textit{cartoon} with the 89\% of them made public. T11: 
\begin{quote}
    \textit{``In a classroom setting, it would be fine not to use the privacy options because I know the people. But if it’s on a platform such as YouTube, I would use the cartoon. ... I prefer the avatar option because I’m shy.''}. 
\end{quote}

Though our participants were instructed to generate and share ASL comments, their primary focus remained on comprehending ASL content, with ASL anonymization considered important but less so in comparison to understanding ASL content. In these cases, \textit{cartoon} met their needs as a balance of both needs. From the interview comments, the rationale behind not sharing, instead of privacy and anonymity concerns, was that some participants believed that certain comments content were less beneficial for peer learners and were more closely tied to their individual learning experiences, serving as personal notes.




\subsubsection{Sharing realistic signmaku increase effort in self-representation} \label{2}The time spent on shared ASL clips (M= 108.8 seconds per person, SD= 91.3) was longer than those set as private (M= 97.7 seconds per person, SD= 200.8) (p= .052), indicating signing shared comments increased users' effort.

Sharing signed comment was also accompanied with the needs in self-representation. Five participants re-recorded their ASL videos, after deciding they would \textit{share} the signmaku with others. 
After including the ASL re-recording, the total time for crafting ASL increased. The total time for crafting ASL comments averaged 168.7 seconds (SD=108.0) per participant, similar to the time for typing text. 
On average, participants had 0.8 re-recorded attempts per participant. 
The interview result explains they re-recorded the ASL clips for better \textit{``precision, clarity and to be more related to the video context''} and \textit{``better lighting and placing in the camera.''} The two participants who favored providing text-only comments (all text comments were shared). They mentioned that text commenting was for the purpose of sharing; typing text allowed them to avoid the stress associated with impression management, expedited the study process, and enabled more precise expression through text editing. And the process of ``editing'' ASL comments may necessitate re-recording attempts, as elaborated in the quotes below: 

\begin{quote}
    \textit{``
    I want to sign of course, but sometimes I feel like I have to set everything up, look nice with the right clothes, have my makeup and hair styled, and all that. Ugh! It seems to take so much time when I could just type it all up and edit it easier and faster...With ASL, I have to think it through to make sure I use the right sign and the correct facial expression.''} -- T4 
\end{quote}

Participants found that using two proposed filters for sharing required less self-representation effort compared to the \textit{realistic} style. After including re-recording, time, the time spent on \textit{realistic} (M= 45.8 seconds per comment, SD= 30.8) was longer than time spent on \textit{robotic} (M=20.5 seconds per comment, SD= 17.1) (p< .05). Some were interested in using two proposed filters for self-presentation, however, the current prototype lacked real-time filtered views, leaving participants uncertain about how well the filters suited them. Participants suggested that having a preview of the filtered video would boost their willingness to use and share with two AI filters by allowing them to check for clear signing and facial movements. We did not conduct further comparison of effort for each type in the context of private vs public options due to our limited sample size.

\subsubsection{Signmaku reduced the cost of creating compared to text-based commenting.} \label{3}
Each participant was encouraged to create at least one text comment and one ASL comment. They had the choice to create comments solely in text, exclusively in ASL, or, on rare occasions, to opt for a combination of both. The prototype tool showed text-only comments anonymously by default, and participants were informed of the options during on-boarding.

Among all the participants, three preferred ASL comments only, and two preferred to create text comments only. We then compared the effort of participants' signmaku and text comments. After dropping five participants who preferred only one commenting method only, the remaining participants (N=15)  uploaded 2.6 \textit{realistic} ASL comments and 2.5 text comments on average. Given the time cost of converting the \textit{realistic} ones to the other two styles and the constraint of each study, we did not have the participants convert their realistic comments to other styles.  Notably, a paired-samples Wilcoxon signed-rank test indicated that the total time length of ASL comments (M= 74.0 seconds, SD= 66.7) was significantly lower than that for typing text comments (M= 190.6 seconds, SD= 153.6) (p< .05). 
Similarly, the average duration for each ASL comment (M= 26.9 seconds, SD= 22.8) was considerably shorter than the typing time for text comments (M= 94.4 seconds, SD= 80.6) (p< .001). 
After including ASL redo, time spent per clip on ASL averaged 61.0 seconds (SD= 19.8), and time spent per person on ASL averaged 168.7 seconds (SD= 108.0). Both were comparable to typing text per clip and per person, respectively (paired-samples Wilcoxon signed-rank test, p>.05). 

In other words, time spent on ASL, including redo, was comparable to the time taken for text comments, both longer than ASL without redo. However, despite the similarity in time between ASL with redo and text comments, participant interviews shed light on the distinct differences: the majority of participants displayed more fluency and expressiveness in ASL than in typing text comments. Furthermore, their preference for ASL over text for language input was evident from their background surveys. A notable statement from T15 encapsulates this sentiment: \textit{``As a Deaf person, fully expressing myself in English is challenging. In ASL, I can do so fluently and effectively''}

The content between text and ASL comments was similar. Both commenting mechanisms were used to share emotions and to relate personal experiences towards the video content as well as peers' signmakus. For example, T8 commented on the video content using text comment: \textit{The use of augmented reality in retail and shopping is intriguing.} They also created a signmaku to reply to peer's signmaku in the same context: \textit{At 4:20 in the video, the student made a comment that I really agree with. I like augmented reality shopping. It's cool that you can pan you phone around the room to see if a couch can be moved around in a room. Once you know it fits, then you can buy it in person} (ASL clip transcribed to English). 

\textbf{Summary-RQ2} 
After participants created their own signmaku comments, we found a positive correlation between participants' perceived physical load when viewing the cartoon signmaku and some questions in perceived social connectedness and enthusiasm for learning. Such a relationship was not found when participants viewed the other two signmaku styles. Additionally, compared to traditional text-based commenting, commenting using ASL costs less time and is more expressive for DHH to create, though they would spare more time on the setup for self-representation, e.g., clothing right, if to share realistic signmaku. 

\section{Discussion}

\subsection{Signmaku as an Edu-tainment Feature Promoting an Inclusive Video-based Learning Community} In our within-subject study exploring the impact of three styles of signmaku (\textit{realistic}, \textit{cartoon}, and \textit{robotic}) on learning, the findings from RQ1 and RQ2 revealed that \textit{cartoon} were preferred over \textit{robotic} and \textit{realistic}. \textit{Cartoon} provided good understandability (RQ1 \& RQ2) while preserving some anonymity (RQ2) which meets their need to share own comments, increases sense of learning community (e.g., inter-dependency) and further elicits the most positive emotion change (RQ2). 

In RQ1, additional remarks highlighted that signmaku, regardless of style, enhances the participation and re-engagement of DHH during the learning process. 
It also imparts a greater wealth of visual information through sign language, a dimension that is absent in written English. Additionally, it assists in alleviating boredom often associated with reading text-only captions. Our discoveries provide a valuable expansion to the current video-based learning experience for DHH, which predominantly concentrates on improving video caption accuracy and designs (as seen in \cite{bhavya2022exploring}). 

RQ1 and RQ2 results suggest that signmaku can enhance inclusivity in video-based learning communities through entertainment and emotional engagement. These findings empirically reinforce the value of signmaku as an ``edu-tainment'' feature, aligning with Buckingham and Scanlon's concept of captivating learners' attention and evoking emotions to promote inclusive video learning \cite{aksakal2015theoretical}. We contribute signmaku as a new ``edu-tainment'' and learner-sourcing feature and reveal its applicability in a minority online learner population- DHH. In this case, these signmaku comments could be viewed as a co-created and sustained outcome stemming from the relationship between DHH and the video. Existing ``edu-tainment'' features in video-based learning include vibrant animations  \cite{aksakal2015theoretical}, creators offering authentic, casual, and personalized content \cite{xia2022millions}, and teachers incorporating entertainment elements on live-streaming platforms \cite{chen2021afraid}.  Earlier studies have explored the use of ``edu-tainment'' features to make sign language learning material more enjoyable for DHH children \cite{rahmalan2018edutainment}.


Our findings show the promise of signmaku in enhancing inclusive video-based learning through the integration of social interaction via a more accessible commenting mechanism and an improved sense of social connectedness. Specifically, our research builds on previous research by introducing DHH learners to a pseudo-synchronous learning experience and allowing them to use their preferred language (sign language) to interact with online learners they do not know, if needed, anonymously. As highlighted in previous research, social interactions have always held importance but have been lacking for DHH individuals during their formative years \cite{stinson1999considerations,constantinou2020technology,seita2018behavioral}. How will signmaku be adopted in the wild? Similar challenges have been identified among other minority learners as well (e.g., \cite{sharma2016teaching}). How may this new commenting feature be adopted by a broader audience? How can it be utilized to promote interaction between learners who have varied accessibility challenges? Our work has paved the way for greater inclusivity in online learning, marking a vast and promising avenue for further exploration and research.



\subsection{Design Implications for DHH-Inclusive Video-based Learning} Our findings demonstrate that ASL offers a more accessible means for individuals who are DHH to both view their peers' comments and make their own comments. Additionally, ASL provides a visually richer source of information compared to text-based comments. Culturally Deaf individuals primarily rely on visual communication channels rather than auditory ones, as supported by previous research \cite{bauman2009reframing, friedner2015s, padden2009inside}. This underscores the significance of accommodating DHH users with ASL and visual learning in video-based learning, paralleling the importance of audio design for hearing users in auditory-focused environments, as demonstrated by previous research. For example, Arakawa et al. developed a video-based learning system that perturbs the voice in the video to help mind-wandering hearing users refocus their attention without consuming their conscious awareness \cite{arakawa2021mindless}. Chen et al. studied audio note-taking in video-based learning tends to be more expressive and lengthy in comparison to text-based note-taking \cite{chen2021exploring}. Our findings provide new design implications centering on how generative AI could be applied for inclusive video-based learning.

\subsubsection{Generative AI for ASL and Text}  The comment made in English could be translated to ASL for DHH learners. Vice versa, ASL comments should be translated into English for general hearing users. With recent advancements in AI, the real-time generation for ASL is possible, e.g., \cite{desai2023asl}. 
It also allows DHH learners to break the circle and communicate with users in another language in their own preferred modality. ASL generated from English is also needed from those DHH who prefer to type or sign only, due to strong English \& ASL skills and ease in editing typed English compared to editing or re-recording signed comments (RQ2). 

Language comprehension is fundamental, and participants further contributed the metric ``smoothness'' for signing avatars, signifying the AI-generated signing avatar's capacity to seamlessly integrate various modalities. When formulating specific metrics for the development of AI-generated ASL content, it is crucial to be included as an evaluation criteria. It's also important to note that previous research has highlighted distinctions in temporal preferences when interacting with human signers versus AI-generated signing avatars. For example, DHH users may require faster signs and slower transitions than those typically employed by human signers \cite{al2021different}. Therefore, while the metrics for human signers can serve as a foundational reference, they may not directly apply to AI-generated signing avatars, necessitating tailored adjustments and refinements.

\subsubsection{The Impact of AI-Generated Facial Expressions on Sign Language Comprehension}
Our findings demonstrate a preference for \textit{realistic} and \textit{cartoon} signmaku over \textit{robotic} signmaku in both viewing, creating, and sharing due to natural hand and body movements and at least some facial expressions making the language possible to understand. 
The \textit{cartoon} signmaku has some facial movement with unfiltered hands, and the task demand was similar to \textit{realistic} in all three dimensions, indicating that when facial cues are lessened, the communication is still understandable. And when hand, body, and facial movements are all lessened, the message becomes hard to understand. The above implies the fundamental importance of hand and body movements. 

Meanwhile, as shown in RQ2, \textit{realistic} is still preferred over \textit{cartoon} for its understandability, especially in sharing. It reveals that facial movement makes it not only understandable but truly expressive. For example, facial expression and body posture are often used to convey emotional magnitude (surprised vs. shocked. vs. aghast) and eyebrows may be raised or lowered depending on the form of question being asked \cite{weast2008questions}. The importance of the expressive body has been brought up by researchers in the accessible computing field in general \cite{spiel2022expressive}. 
Expanding on previous research that focused on how audiences perceive various ASL filtering styles \cite{lee2021american}, our study also considered the perspective of those sharing content. We discovered that users are less inclined to create and share signmakus with filtered styles that lack facial movements.

\subsubsection{The Role of AI-Generated ``Make-Up'' Filters in Enhancing Sharing} RQ2 results suggested that the quality of ASL comments influences participants' williness to share. Some opt for re-recording to enhance quality, and T4 emphasizes the importance of reviewing and improving, particularly in facial movements. Therefore, 'Make-up' filters that automatically improve recording quality could improve willingness to sharing by reducing creation efforts. For example, filters should be able to make signed comments to show more expressive should be more vivid facial and eyebrow movements. Other non-linguistic features could be useful as well, such as lighting improvements.  These filters underscore the potential of generative AI technology in providing understandable and expressive ASL content.  

In RQ1, participants emphasized the desire for self-presentation options, including fun filters, alongside the anonymity feature, consistent with findings in our formative study, RQ2, and prior research \cite{lee2021american}. Future research should explore diverse filter options for facial expressions, body language, and signing styles. 
As RQ2 suggests, sign language is rich in cultural and linguistic diversity, with variations in regional dialects, signing styles, and personalized expressions. Further generative AI technology should allow identity expression within sign language avatars enabling learners to showcase their unique cultural and linguistic backgrounds, preserving the diversity of sign language communities and authenticity of the individuals. This dataset might also be interesting to researchers who want to contrast fluent and non-fluent signing  -- \cite{hassan2022asl}. 
Similar to a study by Arakawa et al. \cite{arakawa2021digital}, which discovered that voice conversion impacted speech ownership and implicit bias in self-presentation, it is essential to consider potential biases introduced by sign language filters.

\subsubsection{Scalability of Signmaku Design Parameters} 
A significant area where AI can make valuable contributions is in automating the fine-tuning of design parameters during round 2 of formative studies (Phase I). Specifically, design aspects like ``duration,'' ``dimension,'' and ``placement in relation to the video content'' can be optimized by harnessing large-scale data and adapting to various user interface layouts. For example, among these parameters, managing the duration can be particularly challenging. We observed that participants comfortably diverted their attention from the learning content for up to 10 seconds and still easily re-assumed the content without feeling stressed. 
Typically, for educational ASL sentences, one full sentence, as determined in our initial formative study, takes approximately 10 seconds. However, it's worth noting that learners might prefer signing for a longer duration, and in our evaluation study (Phase II), our participants signed comments lasted about 30 seconds. In our current approach, we manually edit the videos, but there is potential for designing a tool that offers auto-editing and streamlines this process. Also, our current design only presents one signmaku at one particular time of the video, future research might explore the possibility of displaying multiple signmaku simultaneously.

We want to emphasize that the parameters we selected for each design aspect are solely for our evaluation study (Phase II), which is based on our selected video on AR. We do not claim these parameters are generalized to broader contexts. Importantly, future research should investigate how participants' experiences are impacted when multiple users' comments are presented, as danmaku in nature allows multiple users commenting on the same video scene \cite{chen2017watching, wu2018danmaku}. Additionally, future research would need to evaluate how filters work for some signers but not for others, and how the filters can be further personalized based on how individuals express differently. This is particularly relevant in light of our needfinding, which revealed significant variations in individuals' signing quality and clarity when using filtered signmaku.

\subsubsection{Promoting Deaf Cultural Identity in Online Social Interactions} It's crucial to recognize Deaf as a cultural identity, not as a disability, and promote mutual inclusivity through intercultural dialogue. To foster a sense of belonging within the Deaf culture, future video systems can consider encouraging DHH individuals to teach ASL during interactions with hearing people. Mack et al. found that Deaf signers want to share their language and culture with both hearing family members and friends \cite{mack2020social}, highlighting the need for online learning technology to facilitate intercultural engagement and language learning between Deaf and hearing individuals. Future research should also investigate hearing individuals' perception of signmaku, exploring design features like signmaku frequency control and AI-generated text captions using recent AI technology  \cite{bragg2020exploring}. Additionally, further research is required to understand meaningful two-way translation using AI, moving beyond word-for-word translations.

\subsection{Limitations} 

We acknowledge a few limitations in our study. First, our participants are from a higher education school specifically for DHH students. DHH students from mainstream higher education schools might have different perspectives toward their interactions with other students and technology. Future studies should conduct studies with more DHH students with a more diverse background. 
Second, our study is conducted using American Sign Language (ASL), and the results may be different using other sign languages. Future studies may explore how different sign languages may impact the implementation of signmaku for online learning.
Third, our current signmaku design did not allow participants to customize the parameters of signmaku. During the interview, participants suggested more ways to customize the signmaku experience, such as changing size and location of the signmakus or turn on or off signmakus. Future studies may explore how these interactions may impact the experience and effects of using signmakus in video-based learning. Future research could build beyond our work on understanding sharing effect \textit{after} video watching to \textit{during} video watching.
Fourth, we used facial expression recognition technology to automatically detect participants' emotion while watching the video. We did not use it to examine the detected emotions, rather we only compared the emotion changes. Future studies could apply more advanced tools to capture learners emotions more accurately and re-evaluate our findings. 







\section{Conclusion} 

We introduced a new video commenting feature, Signmaku, which demonstrated promising results in enhancing video-based learning for DHH students. The \textit{cartoon} signmaku created a more engaging and interactive learning environment, positively impacting users' perceived learning engagement and social interaction. Our novel interaction design has the potential to encourage learners to actively participate in the learning process and empower them to express their opinions openly. Our research expands the design space for online learning, thus addressing the specific needs of DHH individuals and providing access to a wide range of educational resources. We emphasize the need to address the limitations of generative AI techniques in creating sign language to ensure accurate and comprehensive content. Adopting signmaku and prioritizing inclusive online learning can unlock significant learning opportunities for DHH students, thereby promoting equity in education and enabling DHH students to thrive alongside their hearing peers.

\begin{acks}
This material is based upon work supported by the National Science Foundation under Grant No. 2119589. Any opinions, findings, and conclusions or recommendations expressed in this material are those of the author(s) and do not necessarily reflect the views of the National Science Foundation. The authors would like to thank the anonymous reviewers for their efforts and valuable comments.  Additionally, we would like to express our gratitude to all the participants for their thoughtful and reflective feedback.
\end{acks}

\bibliographystyle{ACM-Reference-Format}
\bibliography{reference}

\clearpage
\appendix
\onecolumn
\section{Appendices}
\label{appendix}

\subsection{Survey Questions} \label{appendix:survey}

Below, we present the survey questions used in our Phase II evaluation study. Note: In the surveys, we used the term ``ASL-based Danmaku'' instead of ``signmaku.'' The three types of signmakus were also referred to using different terms for easier understanding: realistic signmakus were referred to as ``ASL-only comments;'' cartoon signmakus were referred to as ``Cartoon ASL comments;'' and robotic signmakus were referred to as ``Robot ASL comments.''

\subsubsection{Part 1: Three types of Signmakus for Video-based Learning}
Danmaku is a YouTube-like video-sharing platform. Viewers’ comments appear on the screen and each comment can be customized (e.g., size, color, etc.). The comments go beyond real-time, with previous and later comments being shown together. 

In our tool, you experienced three types of ASL-based Danmaku: ASL-only comments, Cartoon ASL comments, and Robot ASL comments. Rank how much you enjoyed the three ASL-based Danmaku types you saw in the video. \textit{(Tied ranking is not allowed.)}. 

\vspace{2mm}

\textbf{Realistic Signmakus.}

\begin{itemize}

    \item How \textbf{mentally demanding} was \textbf{viewing} the pop-up ASL-only comments? Mentally demanding refers to actions (e.g., thinking, deciding, remembering, looking, searching, etc.) that were hard to do WHILE viewing the ASL-only comments.

    \textit{Choose on a 10-point Likert scale, where 1 is Easy (low demand) and 10 is Hard (high demand).}

    \item How \textbf{physically demanding} was \textbf{viewing} the ASL-only comments? Physically demanding refers to actions (e.g., clicking, pausing, moving body forward to read, switching eye gaze between features, etc.) that were hard to do WHILE viewing the ASL-only comments.

    \textit{Choose on a 10-point Likert scale, where 1 is Easy (low demand) and 10 is Hard (high demand).}

    \item How much \textbf{time pressure} did you feel \textbf{viewing} the ASL-only comments and the video at the same time? Was the pace slow \& relaxed OR rapid \& rushed?

    \textit{Choose on a 10-point Likert scale, where 1 is Slow \& Relaxed and 10 is Rapid \& Rushed.}

\end{itemize}

\textbf{Cartoon Signmakus.}

\begin{itemize}

    \item How \textbf{mentally demanding} was \textbf{viewing} the pop-up Cartoon ASL comments? Mentally demanding refers to actions (e.g., thinking, deciding, remembering, looking, searching, etc.) that were hard to do WHILE viewing the Cartoon ASL comments.

    \textit{Choose on a 10-point Likert scale, where 1 is Easy (low demand) and 10 is Hard (high demand).}

    \item How \textbf{physically demanding} was \textbf{viewing} the Cartoon ASL comments? Physically demanding refers to actions (e.g., clicking, pausing, moving body forward to read, switching eye gaze between features, etc.) that were hard to do WHILE viewing the Cartoon ASL comments.

    \textit{Choose on a 10-point Likert scale, where 1 is Easy (low demand) and 10 is Hard (high demand).}

    \item How much \textbf{time pressure} did you feel \textbf{viewing} the Cartoon ASL comments and the video at the same time? Was the pace slow \& relaxed OR rapid \& rushed?

    \textit{Choose on a 10-point Likert scale, where 1 is Slow \& Relaxed and 10 is Rapid \& Rushed.}

\end{itemize}

\textbf{Robotic Signmakus.}

\begin{itemize}

    \item How \textbf{mentally demanding} was \textbf{viewing} the pop-up Robot ASL comments? Mentally demanding refers to actions (e.g., thinking, deciding, remembering, looking, searching, etc.) that were hard to do WHILE viewing the Robot ASL comments.

    \textit{Choose on a 10-point Likert scale, where 1 is Easy (low demand) and 10 is Hard (high demand).}

    \item How \textbf{physically demanding} was \textbf{viewing} the Robot ASL comments? Physically demanding refers to actions (e.g., clicking, pausing, moving body forward to read, switching eye gaze between features, etc.) that were hard to do WHILE viewing the Robot ASL comments.

    \textit{Choose on a 10-point Likert scale, where 1 is Easy (low demand) and 10 is Hard (high demand).}

    \item How much \textbf{time pressure} did you feel \textbf{viewing} the Robot ASL comments and the video at the same time? Was the pace slow \& relaxed OR rapid \& rushed?

    \textit{Choose on a 10-point Likert scale, where 1 is Slow \& Relaxed and 10 is Rapid \& Rushed.}

\end{itemize}

\subsubsection{Part 2: Perceived Sense of Community for Learning. (20 Questions)}
These 20 questions were developed based on Rovai's Classroom Community Scales \cite{rovai2002development, dawson2006study}. The questions were divided into 2 sets of 10 questions measuring two factors: connectedness and learning. Below, we present these two sets of questions separately. The number before each question represents the order of appearance. For each question, participants may choose between the following options: \textit{Strongly disagree}, \textit{Disagree}, \textit{Neutral}, \textit{Agree}, \textit{Strongly agree}, which is converted to 1-5 in the analysis, and certain statements' answers are converted to 5-1 according to \cite{rovai2002development, dawson2006study}.

\textbf{First}, rank how much you enjoyed providing three ASL-based Danmaku to future learners. \textit{(Tied ranking is not allowed.)} 

\textbf{Then,} answer the following questions. When we refer to ``tool'', please consider you creating and sharing your own ASL comments.

\textbf{Sense of Connectedness}: Compared to using video to learn by myself, 
\begin{oddenumerate}
    \item I feel students using this tool care about each others.
    \item I feel connected to others using this tool.
    \item I do NOT feel a sense of community using this tool.
    \item I feel using this tool is like family.
    \item I feel isolated using this tool.
    \item I trust others using this tool.
    \item I feel I can rely on others using this tool.
    \item I feel like others using this tool depend on me.
    \item I feel uncertain about others using this tool.
    \item I feel confident others will support me using this tool.
\end{oddenumerate}


\textbf{Sense of Learning}: Compared to using video to learn by myself, 
\begin{evenenumerate}
    \item I feel I am encouraged to ask questions using this tool.
    \item I feel it hard to get help when I have a question using this tool.
    \item I feel I receive timely feedback using this tool.
    \item I feel UNeasy that this tool exposes gaps in my understanding.
    \item I feel reluctant to sign/speak openly using this tool.
    \item I feel using this tool results in only modest learning.
    \item I feel others do NOT help me learn using this tool.
    \item I feel I am given plenty opportunities to learn using this tool.
    \item I feel my educational needs are NOT being met using this tool.
    \item I feel that using this tool does NOT promote my desire to learn.
\end{evenenumerate}

\subsubsection{Demographics Information}

\begin{itemize}
    \item What is your status as undergraduate student? 
    
    \textit{Choose one of the following: Freshman; Sophomore; Junior; Senior; Other: [textbox].}
    \item What is your gender identity? 
    
    \textit{[textbox]}
    \item What is your ethnicity? 
    
    \textit{[textbox]}
    \item What is your age? 
    
    \textit{[textbox]}
    \item How do you identify (within the Deaf community)? 
    
    \textit{Choose one of the following: Deaf; Hard of Hearing; Hearing.}
    \item Which of the following statements is true about your immediate family (parents and siblings)? 
    
    \textit{Choose one of the following: One or more of my family members are deaf or hard of hearing; None of my immediate family members is deaf or hard of hearing.}
    \item What language did you use growing up in face-to-face communication? 
    
    \textit{Rank the following options: English; Other SPOKEN language; ASL; Other SIGN language. Participants may choose ``Does not apply'' for any of the options.}
    \item At what age did you learn ASL?
    
    \textit{Choose ``From birth (0-12 months)'' or any number from dropdown box.}
    \item What is your country of origin? 
    
    \textit{[textbox]}
    \item Is English your 1st written language? 
    
    \textit{Choose one of the following: Yes; No.}
    \item Which are you most comfortable with as a classroom instructional language? 
    
    \textit{Choose one of the following: ASL; English; Both ASL and English.}
    \item What is your major (or planned major)? 
    
    \textit{[textbox]}
    
    

\end{itemize}

\end{document}